\documentclass{article}
\usepackage[T1]{fontenc}
\usepackage{authblk}
\usepackage{graphicx}
\usepackage{calc,amsmath,amssymb,amsfonts}
\usepackage[english]{babel}
\usepackage{subcaption}
\usepackage{geometry}
\usepackage{setspace}
\usepackage{comment}
\usepackage{algorithm}
\usepackage{algorithmic}
\usepackage{pdflscape}
\usepackage{pdfpages}
\usepackage{physics}
\usepackage{array}
\usepackage{booktabs}
\usepackage[hidelinks]{hyperref}
\usepackage{csquotes}

\title{Hyperspectral image segmentation with a machine learning model trained using quantum annealer}

\author[1]{Dawid Mazur}
\author[2, 3, 4]{Tomasz Rybotycki}
\author[3, 4]{Piotr Gawron}
\affil[1]{AGH University, al.~Mickiewicza 30, 30-059 Cracow, Poland}
\affil[2]{Systems Research Institute Polish Academy of Sciences, ul.~Newelska 6, 01-447 Warsaw, Poland}
\affil[3]{Nicolaus Copernicus Astronomical Center, Polish Academy of Sciences, ul. Bartycka 18, 00-716 Warsaw, Poland}
\affil[4]{Center of Excellence in Artificial Intelligence, AGH University, al.~Mickiewicza 30, 30-059 Cracow, Poland}

\begin{document}

\maketitle

\begin{abstract}
	Training of machine learning models consumes large amounts of energy. Since
    the energy consumption becomes a major problem in the development and
    implementation of artificial intelligence systems there exists a need to
    investigate the ways to reduce use of the resources by these systems. In
    this work we study how application of quantum annealers could lead to
    reduction of energy cost in training models aiming at pixel-level
    segmentation of hyperspectral images. Following the results of QBM4EO team,
    we propose a classical machine learning model, partially trained using
    quantum annealer, for hyperspectral image segmentation. We show that the
    model trained using quantum annealer is better or at least comparable with
    models trained using alternative algorithms, according to the preselected,
    common metrics. While direct energy use comparison does not make sense at
    the current stage of quantum computing technology development, we believe
    that our work proves that quantum annealing should be considered as a tool
    for training at least some machine learning models.\\
    
    \textbf{Keywords}: RBM, QML, Hyperspectral imaging, image segmentation
\end{abstract}

\section{Introduction}\label{introduction} The rapid growth of artificial
intelligence, especially in the field of generative models~\cite{Goodfellow2014}
and transformer architecture in 2017~\cite{Vaswani2023} has lead to a major
proliferation of large deep learning models. It is becoming a major concern that
economic opportunities that are believed to be existing coming from the
explosion of large models, lead to major energy consumption related to training
and using these models. In order to mitigate this problem it is important to
search for alternative methods of models training. In this work we employ an old
idea and implement it on a new hardware device --- namely a quantum annealer.
The old idea is the Restricted Boltzmann Machine (RBM), initially introduced in
1986 under the name Harmonium~\cite{Smolensky1986}. RBM is a generative model
that has the ability to learn a probabilistic distribution over its set of
inputs. RBM is a widely used machine learning technique for
both unsupervised~\cite{Decelle2023} and supervised tasks \cite{Dixit2021RBM_DWave}.

Among others, RBMs have been used in computer vision, as a part of the
image processing systems. For instance, the authors of QBM4EO devised a
ML model for multi-label land-use classification~\cite{QBM4EO}. Their model
was designed to process hyperspectral data from the Sentinel-2 images dataset,
hence we were able to ascertain the model capabilities to process similar data.
However, instead of multi-label classification, we looked into a different
computer vision task, namely image segmentation.

Segmentation is a core task in computer vision. It's goal is to partition an
image into regions representing different objects or materials. Segmentation
makes it possible to analyze image's structure accurately, and it has
applications in many fields, such as medicine~\cite{Wang2022} or satellite
image analysis~\cite{fajtl2020}. In this work, we focus on the segmentation of
multispectral images. Such images differ from ordinary ones in that they
capture information in broader electromagnetic wavelength spectrum
\cite{Mengu2023}, not just in the visible range. Such data allow for a deeper
analysis of the imaged object properties~\cite{Sifnaios2024}. 

What we strove to investigate was how accurately a model similar to the one
proposed by~\cite{QBM4EO} can handle pixel-level multispectral image
segmentation task. Since the~\cite{QBM4EO} model was trained using quantum
annealers (QA), the problem is essentially a quantum machine learning (QML) one.
QML is an intersection of quantum computing (QC) and machine learning (ML). It
leverages the principles of quantum mechanics to solve complex ML problems. In
our case, the problem is RBM training. Since RBMs were trained far before first
QAs became available, another research question arise. Are quantum training
techniques better than their classical counterparts, at least in this limited
context? If quantum training techniques are at least comparable in terms of
results quality, one can hope that they could find wider applications due to
their ability to use less energy~\cite{Jaschke_Montangero_2023}. 

This paper is organized as follows. We begin with the problem formulation, where
we formally introduce the idea of multi- and hyperspectral images and image
segmentation. Moreover, we discuss previously proposed solutions to the problem
therein. In the next section, we overview quantum machine learning, focusing
especially on the building blocks of the model we propose. Here, we also
describe how quantum annealers can be used to train specific ML models. Then, we
start with describing the experiment. We review the dataset we use and introduce
our model. We then discuss our experiments in great detail and conclude this
section with results analysis. We finish this paper with conclusions, insights
and directions for future research.

\section{Problem formulation}
\label{sec:problem}

Image segmentation is an important area of research and application in the field
of computer vision. It is a process of dividing an image into homogeneous
regions~\cite{Pal1993}. One can find its uses in many fields, such as
medicine~\cite{Ma2024}, remote sensing~\cite{Kemker2018}, and vision systems in
autonomous vehicles \cite{Cakir2022}.

Formally, image segmentation can be defined as follows
\cite{Pal1993}. If $F$ is the set of all pixels and $P: F \to
\{\text{true},\text{false}\}$ is a homogeneity predicate defined on groups of connected pixels, then segmentation is a partitioning of the set $F$ into a set of connected subsets or regions $(S_{1},S_{2},\ldots,S_{n})$ such that:
\begin{equation}
	\label{segmentation_eq}
    \bigcup_{i=1}^{n} S_{i}=F \quad \text{with} \quad S_{i} \cap S_{j} =
    \emptyset, i \neq j.
\end{equation}
\noindent The homogeneity predicate $P(S_{i})=\text{true}$ for all regions
$S_{i}$ and $P(S_{i} \cap S_{j})=\text{false}$, when $S_{i}$ is adjacent to
$S_{j}$. Set $F$ is a digital image defined as 

\begin{equation}
	\label{eq:image}
	F_{W \times H \times B}=[f(x,y,\lambda)]_{W \times H \times B}
\end{equation}
\noindent where $W$ and $H$ are image dimensions, $B$ is the band dimension and
$f(x,y) \in G_{L}=\{0,1, \ldots, L-1\}$ is a set of discrete levels of the
feature value and $(x,y)$ denotes the spatial coordinate. 

In the context of this work, $B$ is of utmost importance, because this
dimension is the one used to determine if an image is multi- or hyperspectral.
Typically, images with 3--15 spectral bands are considered multispectral, 
whereas hyperspectral images can have hundreds of spectral bands
\cite{Hagen2013}. One can clearly see how that makes analysis of such images
more demanding in comparison to the standard ones.

Since image segmentation task is an old and well established problem there are
variety of techniques to deal with it~\cite{Pal1993}. In the context of this
work, the most interesting are the ones using machine learning techniques (ML),
especially unsupervised learning methods. Among those cluster analysis
techniques are popular~\cite{Bratchell1989}. These methods, however, are also
known to have limitations~\cite{GARCIADIAS2020227}. It's therefore reasonable to
explore different options.

The decision of which algorithm to select can be further guided by the
approach one wishes to implement. The image can be analyzed using either
pixel-level or patch approach. The latter usually involves using deep
convolutional neural networks~\cite{Ajit2020, Morchhale2016}. However, research on hyperspectral imaging at the pixel-level has
grown in recent years, leading to an increasing number of scientific
publications in this field~\cite{Morchhale2016, González-Santiago2022, Yan2023,
Sifnaios2024, Chen2024}. We therefore decided to pursue the latter direction.

\section{Machine Learning and Quantum Machine Learning} 
\label{sec:qml}

Artificial intelligence is a technology at the intersection of mathematics and
computer science. It includes a wide spectrum of methods and algorithms that
enable machines to learn, resulting in a wide range of applications
\cite{raschka2019, Goodfellow2016}. A subset of artificial intelligence that
explores machines' capability to learn patterns from the data is called
machine learning (ML).

The focus of this work in on unsupervised learning It is a fundamental approach
in machine learning that allows models to learn patterns in data without
explicit labels~\cite{Naeem2023}. The goal of learning in this context is to
uncover the underlying structure of the dataset. Classic problems in which
unsupervised learning is applied are clustering, dimensionality reduction and
representation learning. Clustering can be performed using a variety of
algorithms~\cite{Bindra2017}, including partitioning methods, density-based and
hierarchical methods. Dimensionality reduction has traditionally been achieved
using principal component analysis (PCA)~\cite{Goodfellow2016}. Lately,
autoencoders have also proven to be highly effective for this
task~\cite{Wang2014}. Representation learning, also called feature learning, is
a process during which algorithms extract meaningful patterns from data to
create representations usually in a lower dimension than the original
data~\cite{Goodfellow2016}.

In the context of this work, unsupervised learning will be used for all the
tasks we mentioned. The model we propose, detailed in section~\ref{sec:exp},
employs both dimensionality reduction and clustering. We also use standard
clustering algorithms as a baseline for out model quality assessment.

\subsection{Latent Bernoulli Autoencoder}\label{lbae_section}

Autoencoders are artificial neural networks used for unsupervised representation
learning. Their architecture consists of two separate networks: encoder and
decoder. Encoder transforms the data into its latent space representation
whereas decoder tries to restore the input data from it. An autoencoder whose
latent space dimension is lower than the input data one is called
undercomplete~\cite{Goodfellow2016}. Learning an undercomplete representation
forces the autoencoder to capture the most significant features of the training
data, thus it essentially implements a representation learning.



Latent Bernoulli Autoencoder (LBAE) --- is a Variational
Autoencoder~\cite{fajtl2020}, that in the context of our work, has a vital
property such that that its latent space is binarized and therefore its output
can be used as an input to a Restricted Boltzmann Machine.

\subsection{Restricted Boltzmann Machines}\label{rbm_section}
Restricted Boltzmann Machines are undirected graphical, energy-based models that
contain a layer of observable variables and a layer of latent variables.
Typically they're referred to as visible and hidden layer,
respectively~\cite{Goodfellow2016, Tieleman2008}. A key feature of RBM is that
there are no connections between neurons in the same layer, hence the
``restricted'' prefix. RBMs are used for tasks such as
classification~\cite{Koziol2014}, feature extraction~\cite{Li2019} and
multispectral image processing~\cite{QBM4EO}. In addition, RBMs are commonly
used as building blocks in other architectures such as Deep Belief
Networks~\cite{Li2019, Goodfellow2016}. RBM training is based on maximizing the
log-likelihood of the training data, and is typically done using gradient-based
techniques. However, parts of the log-likelihood gradient function, such as
so-called negative-phase \cite{Dixit2021RBM_DWave} are hard to compute.
Fortunately, their approximate values can be obtained using the Monte Carlo
Markov Chain methods (MCMC) \cite{Goodfellow2016}. One family of such algorithms
is the contrastive divergence (CD). While widely used, CD relies on limited
Gibbs sampling steps and can get stuck in local minima. 

Alternatively, the negative-phase can be calculated using samples drawn from an
annealer~\cite{Dixit2021RBM_DWave}. The use of annealing can lead to more
efficient sampling than MCMC algorithms. That's because the latter can struggle
to correctly approximate the negative phase of the
gradient~\cite{Goodfellow2016}. Annealing also explores more global
configurations, improving sampling accuracy and learning dynamics in challenging
scenarios.

Simulated annealing (SA) is an optimization algorithm inspired by annealing
process employed in metallurgy. SA treats the optimization problem as a physical
system, with energy representing the objective function value and temperature
controlling the probability of accepting inferior solutions. The algorithm
starts at a high temperature, where energy-increasing movements are allowed, and
then the temperature slowly decreases to find the global minimum of the
objective function~\cite{Benedetti2021}.

Quantum annealing (QA) is an extension of the idea of simulated annealing that
uses quantum effects to search the solution space. Unlike SA, which relies on
classical temperature perturbations, QA uses quantum fluctuations to overcome
energy barriers. Optimization problems, solvable by QA, are most often
represented in terms of an Ising model~\cite{Kadowaki1998}. Although Quantum
Annealing, in principle, follows similar scheme as Adiabatic Quantum
Computing~\cite{Farhi2000}, the difference is that system evolution in QA is not
necessarily adiabatic~\cite{Vinci2017}.

The use of quantum annealing for RBM training represents a novel
approach~\cite{Datta2005, Dixit2021RBM_DWave}. It starts with determining the
probability distribution of the hidden RBM layer for a given set of input data.
The RBM coefficients are then transformed into a Quadratic Unconstrained Binary
Optimization (QUBO) problem. The QUBO is defined by a quadratic function, where
the linear elements correspond to the biases, and the quadratic elements
correspond to the weights between the neurons. The generated QUBO is then
sampled by an annealer, which finds the variables' values corresponding to the
objective function's minimization. In this context, the annealer acts as a
probabilistic sampler, providing samples of the values of hidden and visible RBM
units based on energy minimization. The values of these samples are then used to
update the RBM weights and biases. The algorithm gradually adjusts the model by
comparing the distributions sampled by the annealer with those inferred from the
data, allowing the hidden structures to be correctly represented in the data.
Using the annealer in this context can help more efficient sampling in cases
where gradient methods become insufficient or when the problem has a complex
energy space, making sampling more difficult. At the beginning of the annealing
process, the system exhibits a significantly quantum behavior. As time passes,
and the system cools down, we arrive at the systems final state, which
corresponds to the low-energy solution of the assigned problem. 

\subsection{Cluster Analysis}
Cluster analysis is a category of unsupervised learning algorithms that seek to
divide a given set of objects into homogeneous clusters~\cite{Bratchell1989}. In
the context of image segmentation, cluster analysis divides an image into
regions with similar properties. 

The Agglomerative Hierarchical Clustering (AHC) algorithm relies on a bottom-up
approach. It starts the clustering process by treating each data point as an
individual cluster and then iteratively merging the most similar clusters based
on a selected distance metric. The process continues until all data points are
combined into a single cluster or the desired number of clusters is reached. The
result of such clustering is a hierarchical tree — a dendrogram depicting the
merging of data points into larger and larger clusters~\cite{Bindra2017}.

The AHC  algorithm requires specifying inter- and intra-cluster distance
measures. The latter are commonly called linkage methods, whereas the first are
usually standard distance measures used in analyzed objects processing. In this
work we used Euclidean and Spectral Angle distances~\cite{Sinaice2022}. We also
used complete and average linking criteria~\cite{Müllner2011, Bratchell1989}. In
the context of this work, an interesting property of the AHC algorithm is that
it can be used to further cluster the pre-clustered data. In other words it can
be used to continue initial clustering. This initial clustering could, in
particular, use different algorithm.

The $k$-means algorithm is a clustering algorithm that partition a dataset into
$k$ clusters~\cite{Bindra2017}, aiming to minimize the differences within the
clusters by assigning data points to the nearest clusters. A centroid is the
center of a cluster, a point that represents the average value of all points
assigned to a given cluster. The algorithm begins by initializing centroids,
which represent the centers of the clusters. Next, the distance from each data
point to each centroid is computed, and each point is assigned to the cluster
with the nearest centroid. Once all points have been assigned to their
respective clusters, the centroids are updated by computing the arithmetic mean
of all points assigned to each cluster. This update moves the centroid
positions. Then assigning points to clusters is repeated based on the updated
centroids. The algorithm works iteratively, and the cycle of assigning points to
clusters and updating centroids repeats until the centroids stop changing, which
means that the algorithm converged.

It's worth mentioning that quantum annealing also has the potential to be used
for data clustering. That can be done by transforming the optimization problem
of the clustering cost function into the form of a quadratic binary optimization
problem. QA makes it possible to search the solution space more efficiently and
deal better with local minima compared to classical methods such as
k-means~\cite{Kumar2018}.

\section{Experiments and results}
\label{sec:exp}

We used HyperBlood dataset~\cite{Romaszewski2021} our the experiments. It
consists of fourteen hyperspectral images showing a mock-up crime scene with
bloodstains and other visually similar substances. The images, collected over
three weeks, vary in background composition and lighting conditions. Each image
consists of 120 spectral bands, and is annotated with class labels indicating
the presence of respective substances. This dataset was designed to support the
development and evaluation of machine learning algorithms for hyperspectral
blood detection and classification~\cite{Romaszewski2021}. The dataset consists
of eight classes: background, blood, ketchup, artificial blood, beetroot juice,
poster paint, tomato concentrate, and acrylic paint. Each input datum is a
112-element vector.

\begin{figure}[h!]
    \centering
    {\includegraphics[width=0.44\textwidth]{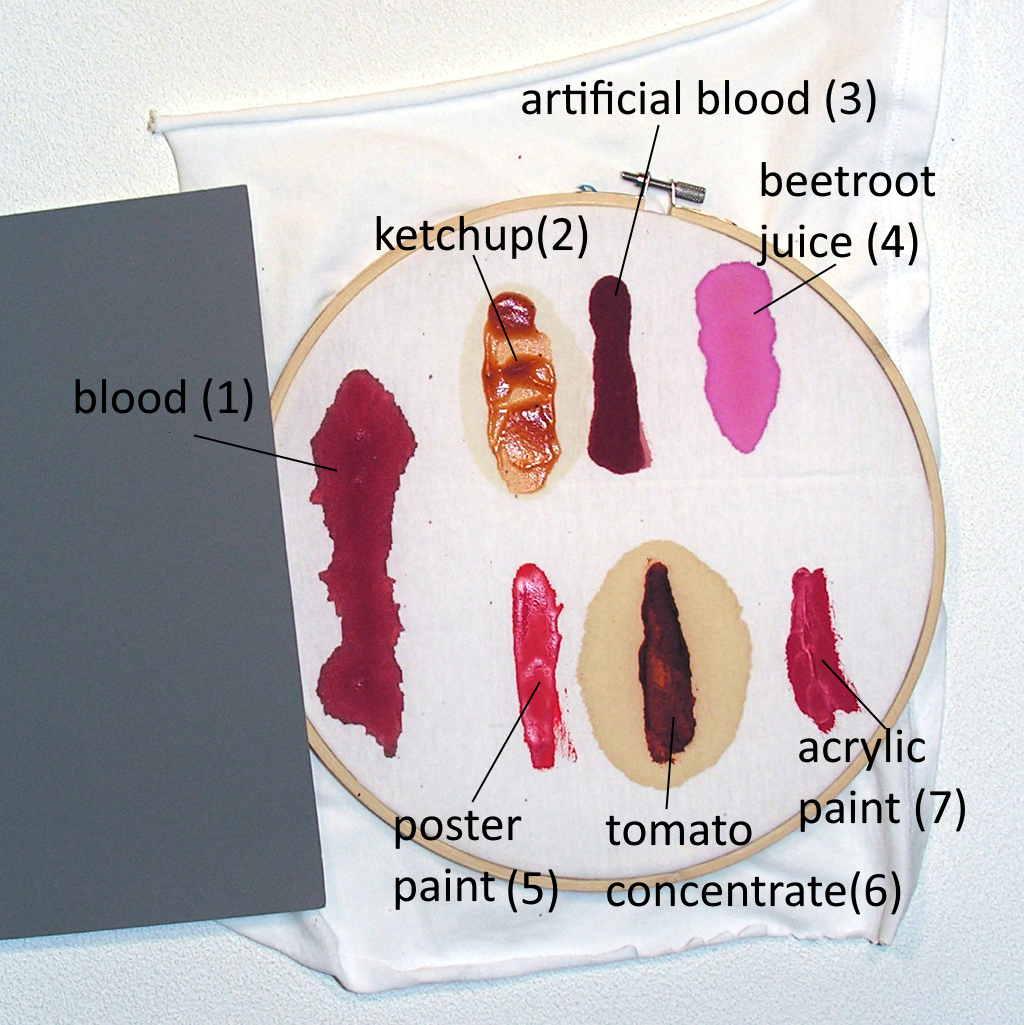}
    \caption[Dataset classes overview.]{An image from the HyperBlood dataset. It shows classes present in the data~\cite{Romaszewski2021}.}
    \label{fig:dataset_img}
    }
\end{figure}

The dataset documentation~\cite{Romaszewski2021} describes some of the
hyperspectral channels as ``noisy''. We decided that the indicated spectral
bands should be removed from further processing. Additionally, we excluded
pixels whose position $(x,y)$ corresponded to the label ``0'' (which is the
background) in the ground truth image. Our goal was to improve homogeneity of
the training data and thus quality of the model.

Further data transformations were to perform normalization, shuffle the data,
and divide the data into training, validation, and test sets. Scaling to the
interval $\langle 0, 1 \rangle$ was chosen as a standard normalization method in
machine learning~\cite{raschka2019} but was also used for processing
hyperspectral images~\cite{Głomb2023}. Data shuffling was used to reduce the
influence of sequential pixel order. Data partitioning was done according to the
Pareto principle: first, the data were divided into a training set and a test
set in a ratio of $0.8:0.2$, and then the training set was further split into a
training and validation set, also in a ratio of $0.8:0.2$~\cite{Murugan2021}.

\subsection{Proposed Model}\label{proposed_model_section} 
\begin{figure}[h!]
    \centering
    \includegraphics[width=1\textwidth]{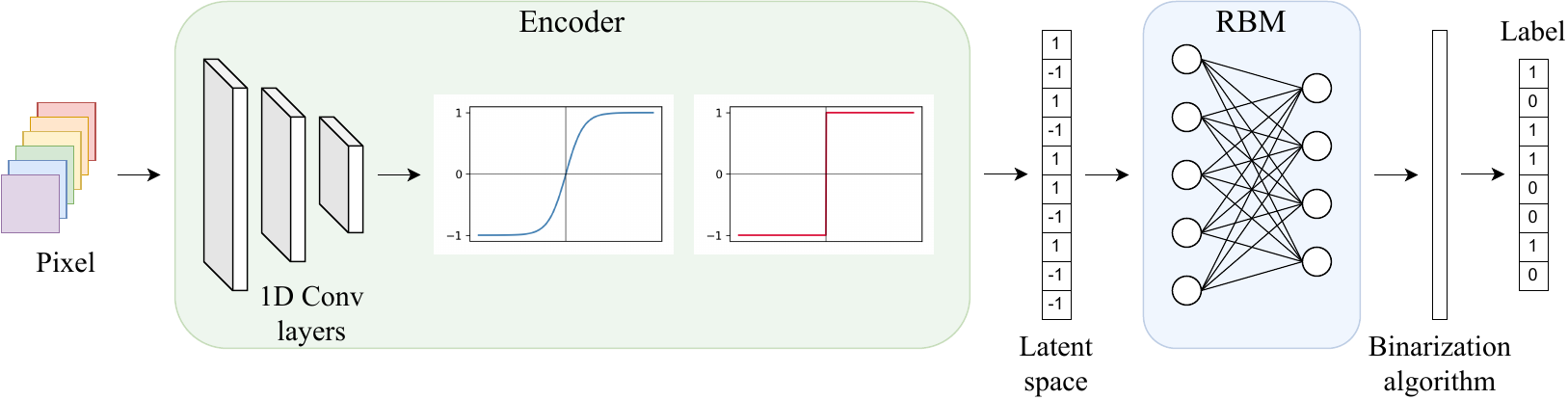}
    \caption[Proposed model pipeline.]{Proposed model pipeline. The pipeline
    	processes spectral pixel data through the LBAE encoder, which consists
    	of one-dimensional convolutional layers followed by a $tanh$ activation
    	and binarization. The resulting binary representation is forwarded into
    	the visible layer of the RBM. Then we compute each hidden layer
    	neuron's activation probability and binarize it. The resulting binary
    	vector is considered the input pixel's label.}
    \label{fig:proposed_model_pipeline}
\end{figure}

The model we propose
consists of two parts. The initial one is encoder of a LBAE. It processes the
input and returns its binarized latent space representation. Binarization is
vital for our model to work, because RBM, which constitutes the second layer of
our model, accepts only binarized inputs. By setting respective neurons of the
RMB visible layer to 1, we update probabilities of neuron activation in the
hidden layer or the RBM. We compute and binarize these probabilities, thus
obtaining a label for the inputted hyperspectral pixel. We present our model in
the figure \ref{fig:proposed_model_pipeline}. Notice that such pipeline is
basically a repurposed QBM4EO pipeline~\cite{QBM4EO}, which was shown to work
well for similar problem.

LBAE is an autoencoder which was introduced in~\cite{fajtl2020} and implemented
in~\cite{QBM4EO}. It contains convolutional layers in both encoding and decoding
parts. Our approach relies on the work done in the~\cite{QBM4EO}. However, our
approach explores the possibility of pixel-level analysis in the context of
hyperspectral images, so we had to adjust~\cite{QBM4EO} implementation of the
LBAE. This adjustment consisted of changing the convolution layers from two- to
one-dimensional. The rest of the network and its layers will stay the same as
in~\cite{QBM4EO} as this project obtained satisfactory performance. LBAE was
trained using standard back-propagation algorithm. Our LBAE transforms the input
pixel data into its binary representation. Due to the undercomplete architecture
of this autoencoder, the dimension of input vector shrinks accordingly. We show
the LBAE convolutional layers parameters in Table~\ref{tab:conv_params}. Each
latent space vector has 28 binary elements.
\begin{table}[ht]
    \centering
    \caption{Parameters of the convolutional layers in the LBAE encoder.}
    \begin{tabular}{@{}lcccc@{}}
    \toprule
    Layer & Padding & Dilation & Kernel Size & Stride \\ \midrule
    Conv1 & 1       & 1        & 3           & 1      \\
    Conv2 & 1       & 1        & 4           & 2      \\
    Conv3 & 1       & 1        & 4           & 2      \\
    Conv4 & 1       & 1        & 3           & 1      \\ \bottomrule
    \end{tabular}
    \label{tab:conv_params}
\end{table}
After the encoding, the data is processed by an RBM. The number of neurons in
the visible layer of our RBM is determined by input data dimension and LBAE
encoder convolutional layers. In the case of our experiments, that would be 28
neurons. 

We trained the RBM with a fixed number of neurons in its visible layer and a
changing number of neurons in its hidden layer $h_i \in \{3, 4, \ldots, 28\}$
using CD-$1$ algorithm. Our approach is to take RBM's hidden layer neurons
activation probabilities and binarize them. The obtained binary vector is a
label assigned to the input pixel. The binarization threshold was selected as
follows: for each $th_i \in \{0.1, 0.2, \ldots, 0.9\}$, we computed the
adjusted Rand score (ARS) between true labels and predicted labels. Then we
checked for which threshold we obtained the highest ARS, and for this threshold,
we compute other metrics. Using the V-measure, we then compare the models. We
selected the $\beta \in [0,1]$ parameter of the V-measure such that the metric
promotes homogeneity --- a metric we deemed more important in the image
segmentation task. For each $i$-th model, we manually tuned $\beta_i \in \{0,
0.01, \ldots, 1\}$, to find such $\beta$ that yield the best V-measure value.
The RBM architecture that most frequently obtained the highest V-measure scores
was selected as the best one. Thus concluding the model architecture design. 

Our approach based on binarized values of neuron activation probability in RBM's
hidden layer leads to a maximum number of unique returned labels equal to $2^N$,
where $N \in \{3,4,\ldots,28$\}. Since there are seven classes in the HyperBlood
dataset after our preprocessing, we see that the RBM will return more labels
than in the dataset. Accordingly, to~\cite{Decelle2023}, RBM may be used to
build relational trees and then use these hierarchies to divide the data into
groups and subgroups. We will use the structure clustered by RBM to create a
distance matrix, which will serve as an input to the Agglomerative Hierarchical
Clustering (AHC) algorithm. By specifying the target number of clusters for the
AHC, we aim to obtain clusterization that will be useful in our segmentation
task. This final phase of the segmentation takes place only for the best model;
after training algorithms are compared.

\subsection{Classical Model Training and Evaluation}
The selection of appropriate evaluation metrics is crucial for an objective
analysis of the quality of the machine learning models. We decided to use the
Homogeneity Score, the Completeness Score, the Rand Score, and the Adjusted Rand
Score, which are commonly used in evaluating clustering
tasks~\cite{Emmons2016},~\cite{Rosenberg2007}. Additionally, Euclidean distance
and Spectral Angle Distance were employed to evaluate the capability of the
autoencoder to reconstruct the data. We trained the model using three different
approaches: the traditional contrastive divergence (CD-$1$) algorithm, an
approach based on simulated annealing (SA) and quantum annealing (QA).

At first, we begin with LBAE training. Since, compared to~\cite{QBM4EO}, we only
changed convolutional layers from two to one-dimensional, we kept the
hyperparameters values. We investigated the influence of changing batch size and
learning rate.

The following values of batch size $b_i \in \{4, 8, 16\}$ and learning rate
$\eta_i \in \{10^{-2}$, $10^{-3}$, $10^{-4}\}$ were chosen as a standard
values~\cite{raschka2019}. For each combination of these parameters, LBAE
training was conducted, ultimately yielding nine trained models. Each model's
performance was evaluated on a test dataset using Euclidean distance and
spectral angle distance (SAD). We used LBAE model that obtained the lowest
values of both metrics, that is the model trained with hyperparameters $b_i=4$
and $\eta_i=10^{-3}$.

Having trained the LBAE model, we determined a baseline for Restricted Boltzmann
Machines (RBM) clusterization that we aim to surpass. For this purpose, we used
the $k$-means algorithm to perform clusterization on the test dataset, on both
raw data and its latent space representation. To avoid the impact of how the
$k$-means centroids are initialized, we conducted the clustering ten times using
different random seeds. Results of the baseline clustering metrics on the test
dataset are included in
Table~\ref{tab:baseline_kmeans_clustering_on_test_dataset}.
\begin{table}
    \centering
    \caption[Comparison of baseline clustering evaluation metrics for k-means and LBAE+k-means.]{Comparison of clustering evaluation metrics for k-means and LBAE+k-means. Values represent the mean $\pm$ standard deviation.}
    \begin{tabular}{@{} lc c @{} }
        \toprule
        \textbf{Metric} & \textbf{$k$-means} & \textbf{LBAE+$k$-means} \\
        \midrule
        \textbf{Homogeneity} & $0.509 \pm 0.020$ & $0.596 \pm 0.056$ \\
        \textbf{Completeness} & $0.443 \pm 0.027$ & $0.520 \pm 0.046$ \\
        \textbf{ARS} & $0.362 \pm 0.065$ & $0.395 \pm 0.090$ \\
        \textbf{Rand Score} & $0.779 \pm 0.021$ & $0.789 \pm 0.031$ \\
        \bottomrule
    \end{tabular}
    \label{tab:baseline_kmeans_clustering_on_test_dataset}
\end{table}

\begin{figure}[h]
    \centering
    {\includegraphics[scale=0.7]{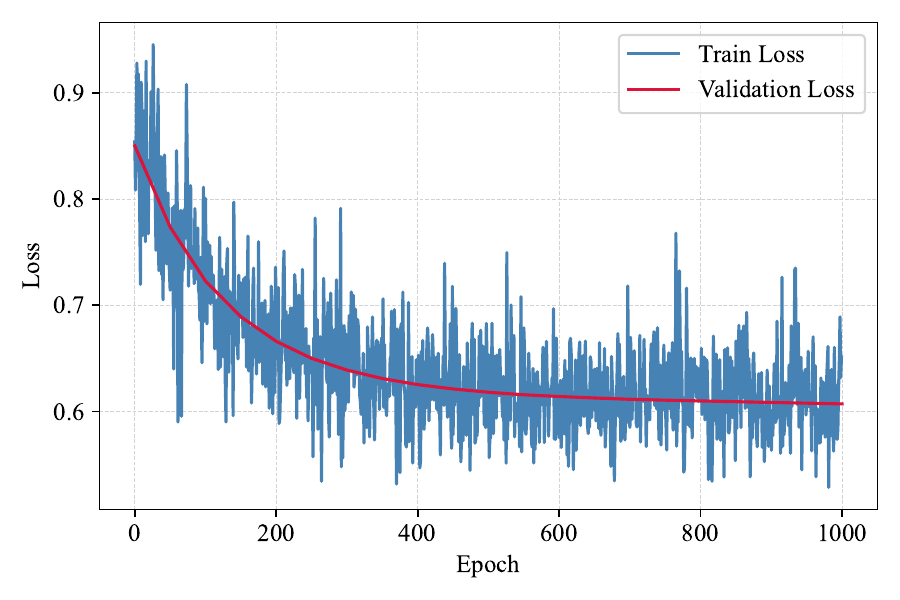}}
    \caption[Learning curve of the RBM from the CD-$1$ training.]{CD-1 trainig learning curve of the RMB with 23 neurons in the hidden layer. The blue line represents the loss value on the training dataset, and the red line represents the loss value on the validation dataset.}
    \label{fig:rbm_learning_cd}
\end{figure}

We then proceeded with the RBM training using CD-$1$. Similar to the case of
LBAE, we keep hyperparameters as they were in~\cite{QBM4EO}, except for one of
them --- a number of neurons in the hidden layer. RBM training experiments were
conducted for the following number of neurons in hidden layer $h_i \in \{3, 4,
\ldots, 28\}$, and those experiments were repeated ten times for different
weights initialization. This experiment concluded that the RBM model with 23
neurons in the hidden layer was the most promising for the segmentation task.
Figure~\ref{fig:rbm_learning_cd} shows the learning curve for that model.
We noticed that our model is returning more unique labels than the target number
of labels on ground-truth images. Following the idea that the RBM returned
structure is hierarchical~\cite{Decelle2023}, we could pass this structure to
another algorithm, known as Agglomerative Hierarchical Clustering (AHC), and
specify the target number of clusters.

We want to compare the final segmentation with a reliable baseline. We, again,
used the standard $k$-means algorithm to obtain it.
Figure~\ref{fig:kmeans_segmentation} illustrates the pixels clustering using
$k$-means, and Figure~\ref{fig:ground_truth_image} illustrates the ground truth
image. Table~\ref{tab:segmentation_metrics_comparison} shows metrics comparison
for created segmentation images. Then, we created a segmentation images using
our model. The results are presented in
Figure~\ref{fig:cd1_rbm_ahc_complete_seg} and
Figure~\ref{fig:cd1_rbm_ahc_average_seg}.

\subsection{Quantum Model Training and Evaluation}
\begin{figure}[h]
    \centering
    \begin{subfigure}{0.49\linewidth}
    {\includegraphics[width=\textwidth]{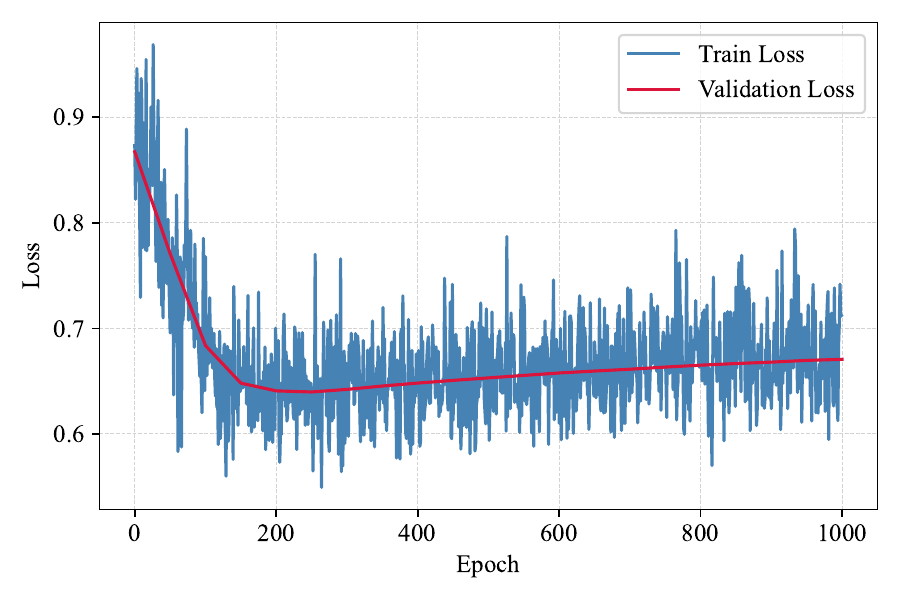}
    	\caption{Simulated annealing.}
    	\label{fig:sim_annealing}
    }
	\end{subfigure}
	\begin{subfigure}{0.49\linewidth}
		{\includegraphics[width=\textwidth]{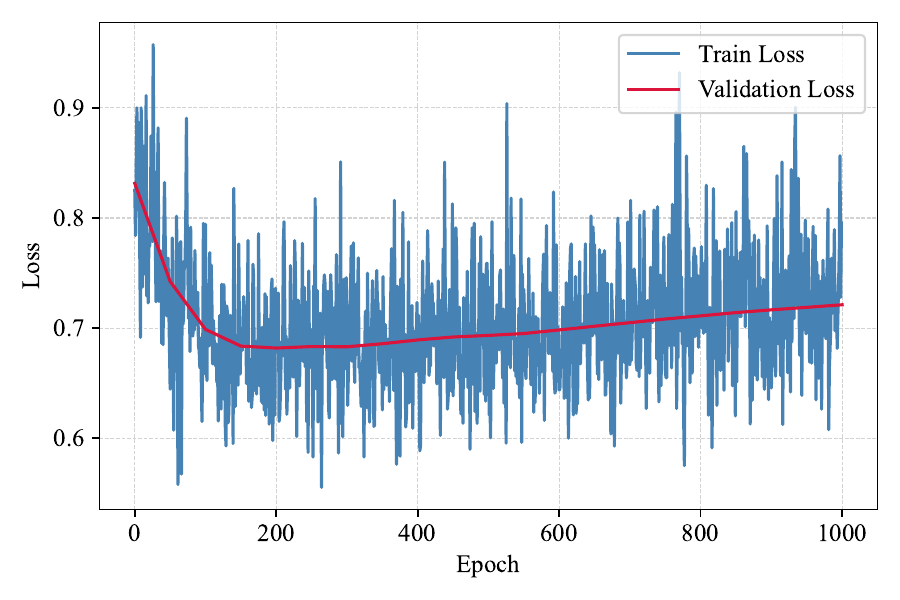}
			\caption{Quantum annealing.}
			\label{fig:quantum_annealing}
		}
	\end{subfigure}
	\caption{Learning curve of the RBM with respective algorithms. The blue
	line represents the loss value on the training dataset, and the red line
	represents the loss value on the validation dataset.}
	\label{fig:rbm_training}
\end{figure}

The next step of the project was to test the implemented training algorithm
using the annealer-based algorithms. First, we used D-Wave's implementation of a
simulated annealing sampler~\cite{DWaveDocumentation}. The training was repeated
ten times for initialization with different weights, this time only for an RBM
model containing 23 neurons in the hidden layer. We decided to save a model
after every other hundred training epochs to execute an insightful analysis of
model performance in the context of clusterization metrics. The best model,
according to the V-measure, was the one after 200 training epochs. The learning
curve of the RBM trained using the simulated annealer is shown in
Figure~\ref{fig:sim_annealing}.

Next, we trained the model using quantum annealing. Again, we used the sampler
provided by D-Wave~\cite{DWaveDocumentation}. We also used the automatic
embedding of our problem into the target QPU --- D-Wave Advantage 5.4 system.
The training was repeated ten times for initialization with different weights,
again only for an RBM model containing 23 neurons in the hidden layer. We
decided to save a model after every other hundred training epochs. The best
model, according to V-measure, was obtained after 300 training epochs. The
learning curve of the RBM trained using the quantum annealer is shown in
Figure~\ref{fig:quantum_annealing}.

\subsection{Segmentation Results}
\begin{figure}[h]
    \centering
    \begin{subfigure}{0.475\textwidth}
        \includegraphics[width=\textwidth]{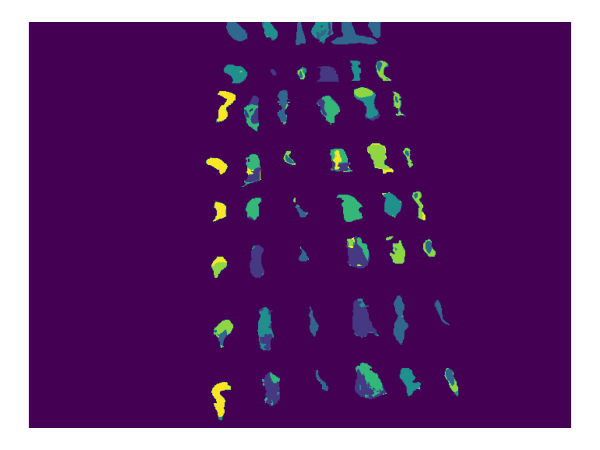}
        \caption{$k$-means segmentation.}
        \label{fig:kmeans_segmentation}
    \end{subfigure}
    \hfill
    \begin{subfigure}{0.475\textwidth}
        \includegraphics[width=\textwidth]{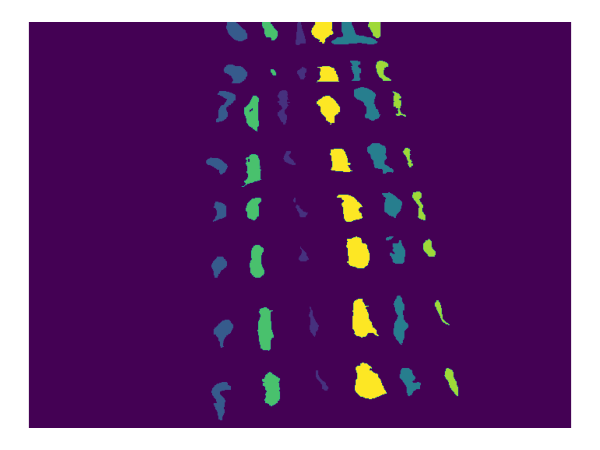}
        \caption{Ground truth.}
        \label{fig:ground_truth_image}
    \end{subfigure}

    \caption{Baseline segmentation by $k$-means and ground truth.}
    \label{fig:baseline_and_gt}
\end{figure}

\begin{figure}
    \centering

    \begin{subfigure}{0.475\textwidth}
        \includegraphics[width=\textwidth]{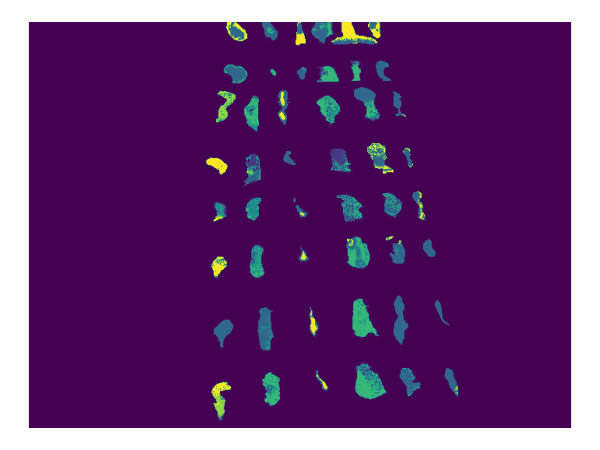}
        \caption{Segmentation by RBM trained by CD-$1$ with AHC using complete linkage.}
        \label{fig:cd1_rbm_ahc_complete_seg}
    \end{subfigure}
    \hfill
    \begin{subfigure}{0.475\textwidth}
        \includegraphics[width=\textwidth]{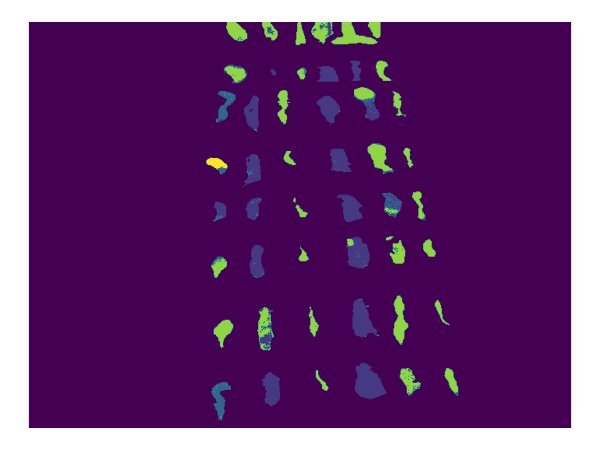}
        \caption{Segmentation by RBM trained by CD-$1$ with AHC using average linkage.}
        \label{fig:cd1_rbm_ahc_average_seg}
    \end{subfigure}

    \begin{subfigure}{0.475\textwidth}
        \includegraphics[width=\textwidth]{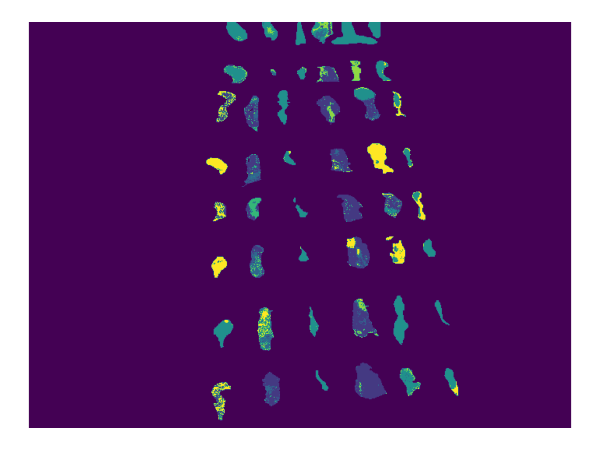}
        \caption{Segmentation by RBM trained by SA with AHC using complete linkage.}
        \label{fig:sa_rbm_ahc_complete_seg}
    \end{subfigure}
    \hfill
    \begin{subfigure}{0.475\textwidth}
        \includegraphics[width=\textwidth]{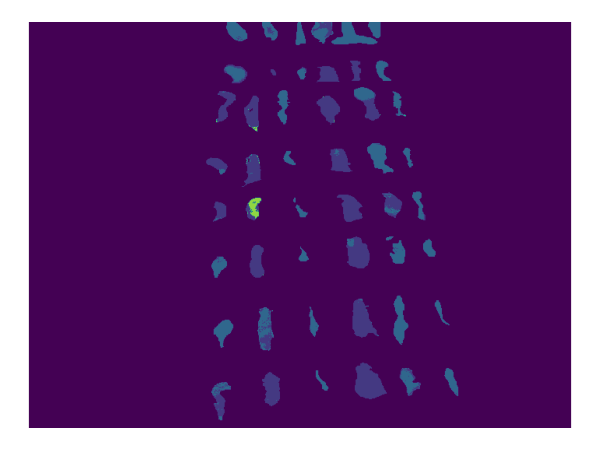}
        \caption{Segmentation by RBM trained by SA with AHC using complete linkage.}
        \label{fig:sa_rbm_ahc_average_seg}
    \end{subfigure}

    \begin{subfigure}{0.475\textwidth}
        \includegraphics[width=\textwidth]{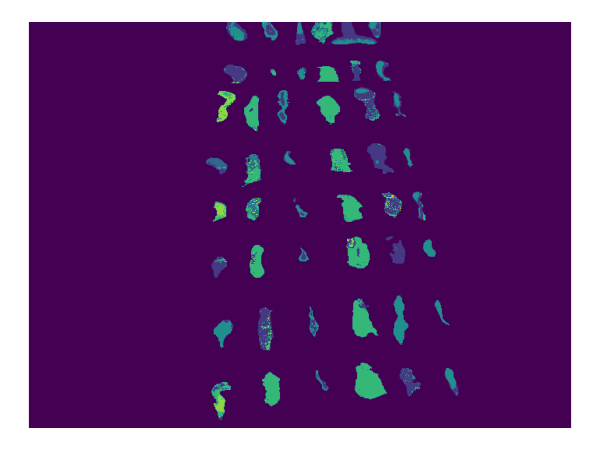}
        \caption{Segmentation by RBM trained by QA with AHC using complete linkage.}
        \label{fig:qa_rbm_ahc_complete_seg}
    \end{subfigure}
    \hfill
    \begin{subfigure}{0.475\textwidth}
        \includegraphics[width=\textwidth]{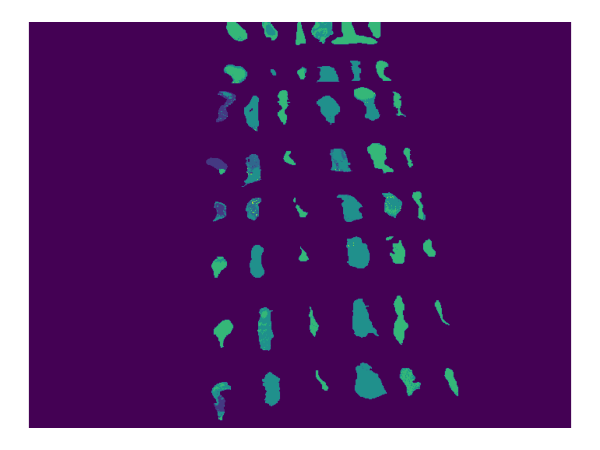}
        \caption{Segmentation by RBM trained by QA with AHC using complete linkage.}
        \label{fig:qa_rbm_ahc_average_seg}
    \end{subfigure}
    
    \caption{Comparison of segmentation results.}
    \label{fig:comparison_seg}
\end{figure}

We present the comparison our experiments results with our baseline segmentation
obtained by using $k$-means algorithm on the
Figure~\ref{fig:kmeans_segmentation} and ground truth image on the
Figure~\ref{fig:ground_truth_image}. They are followed by the segmentations
obtained using our model trained with contrastive divergence (CD-$1$) finalized
with AHC using both linkage methods --- complete and average, those will be
respectively Figure~\ref{fig:cd1_rbm_ahc_complete_seg} and
Figure~\ref{fig:cd1_rbm_ahc_average_seg}. Similarly, we present segmentations
for models trained with simulated annealing (SA) and quantum annealing (QA). For
simulated annealing obtained images are Figure~\ref{fig:sa_rbm_ahc_complete_seg}
and Figure~\ref{fig:sa_rbm_ahc_average_seg}, and for quantum annealing obtained
images are Figure~\ref{fig:qa_rbm_ahc_complete_seg} and
Figure~\ref{fig:qa_rbm_ahc_average_seg}.

A summary of the computed evaluation metrics such as homogeneity, completeness,
adjusted Rand score (ARS) and Rand score (RS) for each segmentation obtained is
presented collectively in Table~\ref{tab:segmentation_metrics_comparison}.
\begin{table}
    \centering
    \caption[Comparison of clustering evaluation metrics for baseline segmentation and segmentation produced by our models.]{Comparison of clustering evaluation metrics for baseline segmentation and segmentation produced by our models. Here, RBM+AHC-C denotes the RBM with AHC using complete linkage, and RBM+AHC-C denotes the RBM with AHC using average linkage.}
    \begin{tabular}{@{} llcccc @{}}
        \toprule
        \textbf{Metric} & \textbf{Training} & \textbf{$k$-means} & \textbf{RBM} & \textbf{RBM+AHC-C} & \textbf{RBM+AHC-A} \\
        \midrule
        \textbf{Homogeneity} & -- & 0.368 & -- & -- & -- \\
                             & CD-$1$ & -- & 0.492 & 0.232 & 0.243 \\
                             & SA  & -- & \textbf{0.510} & 0.231 & 0.171 \\
                             & QA  & -- & 0.505 & 0.254 & 0.260 \\
        \midrule
        \textbf{Completeness} & -- & 0.354 & -- & -- & -- \\
                              & CD-$1$ & -- & 0.239 & 0.257 & 0.416 \\
                              & SA  & -- & 0.244 & 0.266 & 0.381 \\
                              & QA  & -- & 0.282 & 0.319 & \textbf{0.445} \\
        \midrule
        \textbf{ARS} & -- & 0.244 & -- & -- & -- \\
                     & CD-$1$ & -- & 0.163 & 0.129 & 0.242 \\
                     & SA  & -- & 0.180 & 0.199 & 0.181 \\
                     & QA  & -- & \textbf{0.308} & 0.284 & 0.246 \\
        \midrule
        \textbf{Rand Score} & -- & 0.769 & -- & -- & -- \\
                            & CD-$1$ & -- & 0.793 & 0.690 & 0.660 \\
                            & SA  & -- & \textbf{0.799} & 0.711 & 0.590 \\
                            & QA  & -- & 0.790 & 0.729 & 0.656 \\
        \bottomrule
    \end{tabular}
    \label{tab:segmentation_metrics_comparison}
\end{table}

Analyzing obtained metrics on the segmentation images, we notice that almost all
metrics improve their value for RBM trained with QA, and if they are not
improved, they are close to the best-obtained value. For a model that does not
use AHC, we note the consistent improvement of the metrics while changing the
RBM training algorithm from CD-1 through SA to QA. For the metrics homogeneity
and ARS, the RBM model trained with QA surpassed the baseline value, which was
set by the $k$-means algorithm. The use of the AHC algorithm in combination with
RBM affects clustering characteristics. First, AHC improves the completeness
metric, especially when using the RBM version trained with Quantum Annealing. At
the same time, adding AHC reduces the clusters' homogeneity. 

\section{Conclusions}
We proposed a hybrid neural network architecture consisting of Latent Bernoulli
Autoencoder encoder connected with Restricted Boltzmann Machines for
hyperspectral image segmentation. Firstly, the LBAE's encoder part is used for
dimensionality reduction and spectral data binary representation encoding,
necessary for RBM processing. Secondly, we use RBM to perform clustering on the
encoded spectral pixels.

The proposed architecture and experiments confirmed the effectiveness of applying quantum annealing techniques and RBM models training. One can clearly
see that since we achieved better segmentation results, in terms of observed metrics, than the baseline segmentation obtained using standard algorithm --- $k$-means. The results suggest the potential for further research on using quantum annealing in the image processing field.

\section*{Acknowledgments}\label{acknowledgments}
\addcontentsline{toc}{chapter}{Acknowledgments}
\sloppy
We gratefully acknowledge Poland's high-performance Infrastructure PLGrid ACK Cyfronet AGH for providing computer facilities and support within computational grant no. PLG/2024/017155 through and no. PLG/2024/017550.

The authors would like to thank Etos sp.\ z\ o.o.\ and the QBM4EO team for
providing the QBM4EO code. We gratefully acknowledge the funding support by
program ``Excellence Initiative — Research University'' for the AGH University
of Kraków as well as the ARTIQ project ARTIQ/0004/2021.

\bibliographystyle{plain} 
\bibliography{bibliography}

\end{document}